\documentclass{article}

\usepackage{graphicx}
\usepackage{latexsym}
\usepackage[fleqn]{amsmath}
\usepackage{amssymb}
\usepackage{amsbsy}
\usepackage{nishimura-pp}

\title{Stabilization by Unbounded-Variation Noises\thanks{This work was partially supported by Grant-in-Aid for Young Scientists (B) of KAKENHI (25820184)). E-mail: yunishi@yamaguchi-u.ac.jp}}

\author{Y\^uki Nishimura\footnotemark[2]}

\begin{document}

\renewcommand{\thefootnote}{\fnsymbol{footnote}}

\footnotetext[2]{Kagoshima University}

\renewcommand{\thefootnote}{\arabic{footnote}}

\hyphenation{Lyapunov}
\hyphenation{Markov}
\hyphenation{Alcaraz}
\hyphenation{Kushner}
\hyphenation{Krasovskii}
\hyphenation{Schultz}
\hyphenation{Zubov}
\hyphenation{Vannelli}
\hyphenation{Vidyasagar}
\hyphenation{Bucy}
\hyphenation{Mao}
\hyphenation{Hamiltonian}
\hyphenation{Lagrangian}
\hyphenation{Hess}
\hyphenation{Euclid}
\hyphenation{Poisson}
\hyphenation{Wiener}
\hyphenation{Nishimura}
\hyphenation{Yamashita}
\hyphenation{Gaussian}
\hyphenation{Brockett}

\maketitle

\begin{abstract}                          % Abstract of not more than 200 words.
%Asymptotic stability is an important property for analyzing the motion of dynamical systems. 
%In control theory, the property is also important because it ensures that the trajectories of the systems converges on the desired steady-states. The property is generally stated by the context of Lyapunov stability theory, and is developed to control strategies by using control Lyapunov functions. 
%However, asymptotic stability for dynamical systems represented by ordinary differential equations is distinctly different from asymptotic stability in probability for dynamical systems represented by stochastic differential equations. 
In this paper, we claim the availability of deterministic noises for stabilization of the origins of dynamical systems, provided that the noises have unbounded variations. To achieve the result, we first consider the system representations based on rough path analysis; then, we provide the notion of asymptotic stability in roughness to analyze the stability for the systems. In the procedure, we also confirm that the system representations include stochastic differential equations; we also found that asymptotic stability in roughness is the same property as uniform almost sure asymptotic stability provided by Bardi and Cesaroni. 
After the discussion, we confirm that there is a case that deterministic noises are capable of making the origin become asymptotically stable in roughness while stochastic noises do not achieve the same stabilization results.
\end{abstract}

\section{Introduction}

Asymptotic stability (AS) is an important property for analyzing the motion of dynamical systems. In control theory, the property ensures that trajectories of systems converges on the desired steady-states \cite{haddad,khalil,sontag1989}. In a particular case, the convergence is achieved by just adding white noises; the strategy that directly aims at this phenomena is said to be {\it stabilization by noise} \cite{appleby2008,arnold1990,mao1994,nishimura2013cdc}.

The strategy of stabilization by noise provides the origins with {\it asymptotically stable in probability}; however, the stability is weaker than AS because it does not ensure the existence of invariant sets, which are proper closed subsets of Euclidean spaces. The stochastic stability property ``almost the same'' as AS is uniform almost sure asymptotic stability (UASAS) \cite{bardi2005,cesaroni2006}; all the sublevel sets of the related Lyapunov functions are invariant sets with probability one. However, achieving the property is generally difficult because the necessary and sufficient conditions are restrictive. Furthermore, there is a negative result that the non-AS-origin never becomes the UASAS-origin by the addition of any diffusion term \cite{nishimura2013sice}; that is, as long as we employ the stabilization by noise  with the use of Wiener processes, we should allow the possibility of all sample paths traveling to points very far away from the origin.

On the other hand, we claim that the stabilization by noise has still possibility to provide the non-AS origins with AS, provided that the resulting systems are neither ordinary nor stochastic differential equations. To achieve the result, we consider the addition of deterministic noises having unbounded variations. This plan needs to employ a system representation by using rough path analysis \cite{friz2014,lyons1995,lyons2002,lyons2007} because the theory enable us to consider the dynamics including signals having unbounded variations  such as Wiener processes  and a particular kind of deterministic processes. The key point of the analysis is to classify external inputs by calculating the orders of the variations that is finite. Based on the information of the orders, rough paths and their dynamics---generally said to be rough differential equations, and rough systems in this paper---are obtained.

%The unification of ordinary and stochastic differential equations is achieved inspired by {\it the approximation theorems}, which are construction procedures of continuous-time Wiener processes by interpolating discrete-time Wiener processes with piecewise continuous functions \cite{mcshane1972,sussmann1991,wong1965} (see also \cite{nishimura2013cdc}). 
%%In addition, comparing the previous works of the approximation algorithm based on the approximation theorems \cite{liu1997}, the rough path analysis provides rough systems as not an ``approximation'' of the original dynamical systems; that is, rough systems are one of essential system representations. %having mathematical validity as with ordinary and stochastic differential equations. %In addition, the analysis enable us to consider stochastic systems influenced by various stochastic noises such as non-semimartingale processes or delayed Wiener processes \cite{lyons2007}. Thus, rough path analysis has possibility to make great strides of nonlinear system control theory because of its expressive ability.
%Consequently, it should be expected that deterministic and stochastic stability properties are unified on rough systems; and further, deterministic signals having unbounded variations can be used for the strategy of stabilization by noise based on rough path analysis.

In this paper, we  provide a concrete system representation of rough systems in \cite{nishimura2015micnon}, show that {\it asymptotic stability in roughness (ASiR)}---this is the property for the origins of rough systems---is the same property as UASAS, and demonstrate that ASiR is available for stabilization strategies by using deterministic signals having unbounded variations. Especially, we confirm that there is a case that such deterministic noises are capable of stabilizing the origin of dynamical systems in the sense of ASiR while stochastic  noises such as Wiener processes do not achieve the origin being UASAS. 
%Furthermore, we compare rough systems with approximate systems based on the approximation algorithm to clarify the difference of two systems.

The rest of this paper is organized as follows. In \sref{sec:motivation}, we briefly state the motivation of our research by showing the strategy of stabilization by noise using stochastic and deterministic control inputs. In \sref{sec:roughpath}, we summarize the previous results of rough path analysis; especially we show the definitions of geometric rough paths, integration along rough paths, and rough systems. Then, \sref{sec:main} shows main results of this paper; in \ssref{subsec:canonical}, we derive a concrete system representation for our purpose by using canonical rough paths; in \ssref{subsec:stability}, we provide ASiR property; in \ssref{subsec:relation}, we confirm that the ASiR is the same property as UASAS; furthermore, in \ssref{subsec:case}, we illustrate the answer to our motivational example and discuss a further advantage of the usage of deterministic noises against stochastic ones. \sref{sec:conclusion} concludes this paper with some important remarks for control problems.

%%%%%
%\subsubsection*{Notations.}
{\it Notations.}
%Throughout the paper, we use the following notations. 
Let $\R^n$ denote an $n$-dimensional Euclidean space; in particular, $\R$ denote $\R^1$. 
%For integers $a$ and $b > a$, $\N_a^b := \{a,a+1,\ldots,b\}$, especially $\N_a^\infty := \{a,a+1,\ldots\}$. 
If $\gamma : [0,\infty) \rightarrow [0,\infty)$ satisfies $\gamma \in K_\infty$, then $\gamma(r)$ is monotone increasing, $\gamma(0)\equiv 0$, and $\gamma(r) \rightarrow \infty$ as $r \rightarrow \infty$. 
%The probability and the expectation of some event $A$ are written as $\pr \{ A \}$ and $\ex \{A \}$, respectively. 
Further, $w_t \in \R^d$ is a $d$-dimensional, independent, and standard Wiener process; i.e., the values of the variances are all $t$, and all the covariances are constantly zero. The differential forms of It\^o and Stratonovich integrals of $\sigma: \R^n \rightarrow \R^d$ in $w_t \in \R^d$ are denoted by $\sigma(x) dw$ and $\sigma(x) \circ dw$, respectively. For $v:\R^n \to \R$, $g_1,g_2:\R^n \to \R^n$ and $x \in \R^n$, we use the notation of $(\lie{g_1}{v})(x)=(\partial{v}/\partial {x})(x)g_1(x)$ and $(\lie{g_2}{\lie{g_1}{v}})(x)=(\partial{\lie{g_1}{v}}/\partial{x})(x) g_2(x)$.

For the usage of rough path analysis, let  $J=[0,T]$ with $T>0$; a subdivision of $J$ is an increasing sequence of real numbers $D=(t_0,t_1,\ldots,t_N)$ with $N=1,2,\ldots$ such that $0\le t_0 < t_1 <\cdots < t_N \le T$; and $|D_T|=\max\{ t_0, t_1-t_0, \ldots, t_N-t_{N-1}, T-T_N\}$ is a maximum value of lengths of time subintervals for some fixed $D$; further,
\begin{align}
&\Delta_T := \{(s,t) \in [0,T]^2 | 0 \le s \le t \le T \}
\end{align}
be a set of a pair of the initial and the terminal values for time variables; and
%\begin{align}
%&D_T:= \{(t_{k-1},t_k) \in \Delta_T| s = t_0 < \cdots < t_N=t,\nonumber \\
%&\hspace{2.5cm} k =0,1,\ldots,N;\ N =1,2,\ldots\},\label{eq:DT}
%\end{align}
%denote a set of a pair of the initial and the terminal values for time variables of time sub-intervals;}
$\Delta_\infty$ is $\Delta_T$ with $T \rightarrow \infty$. Let also
\begin{align}
T^{2}(\R^n)=\R \oplus \R^n \oplus (\R^n \otimes \R^n);
\end{align}
that is, an element of $T^{2}(\R^n)$ consists of a scalar, a vector, and a matrix.
%\begin{align}
%&T^{(\lambda)}(\R^n)=T((\R^n))/B_\lambda, \\
%&T((\R^n)):=\{a=(a_0,a_1,\ldots)| a_\lambda \in (\R^n)^{\otimes \lambda}\  \forall \lambda \in \N_0^\infty \},\\
%&B_\lambda := \{a=(a_0,a_1,\ldots)|a_0=\cdots=a_\lambda=0\},\ \lambda \in \N_0^\infty,
%\end{align}
%and $\lfloor p \rfloor$ be the integer satisfying $p - \lfloor p \rfloor \in [0,1)$ for $p > 0$, where
%\begin{align}
%(\R^n)^{\otimes \lambda} := \underbrace{\R^n \otimes \cdots \otimes \R^n}_{\lambda}
%\end{align}
%Further, $a \otimes b$ denote a tensor product if $a$ and $b$ are tensors. 
For any $ A, B\in T^{2}(\R^n)$ with $n=1,2,\ldots$ and $\lambda =0,1,2$, a product of $A$ and $B$ is defined by $A \otimes B = C \in T^{2}(\R^n)$, where
\begin{align}
C^\lambda := \sum_{l=0}^{\lambda} A^{l} \otimes B^{\lambda-l}.
\end{align}
For any spaces $M$ and $N$, the coordinate projections are defined by $\pi_{M}: M \oplus N \rightarrow M$ and $\pi_{N}: M \oplus N \rightarrow N$. For any function $y_\tau: [0,\infty) \rightarrow Y$ for any set $Y$, $y_{s,t}: \Delta_T \rightarrow Y$ is defined by $y_{s,t}=y_t-y_s$. The $i$-th element of a vector $a \in \R^d$ is represented by $a[i]$, and the $i$-th row and the $j$-th column element of a matrix $A \in \R^d \otimes \R^d$ is denoted by $A[i,j]$. For $c \in \R$, $b^c$ means $b$ to the power of $c$ if $b \in \R$, and $B^c$ denotes the $c$-th element of $B$ if $B \in T^{2}(\R^n)$ (see also \dref{def:multiplicative}).%; $y^{Letters}$ is mere a mathematical symbol if $Letters$ starts a large letter.

%%%%%%%%%%
\section{Motivational Example}\label{sec:motivation}

\begin{figure}[tbp]
%\centerline{\includegraphics[width=7cm]{sussmann_w11.eps}}
\centerline{\includegraphics[width=6cm]{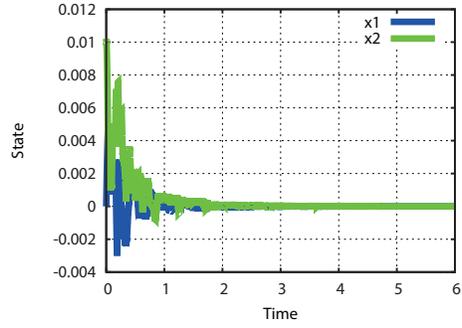}}
\caption{A sample path of \eqref{eq:simple1} with $u=\dot{w}^{Mc}$.}
\label{fig:mcawp-state}
\end{figure}

To begin with, let us consider a stabilization problem for the origin of the following simple system:
\begin{align}\label{eq:simple1}
&\dot{x}_t= f(x_t) + g_1(x_t)u_t[1] + g_2(x_T)u_t[2],
\end{align}
where
\begin{align}
&f(x)=\begin{bmatrix}
-7 & 0 \\ 0 & 1
\end{bmatrix} x, \ g_1(x)= \begin{bmatrix}
0 & 0 \\ 1 & 0
\end{bmatrix} x, \ g_2(x)= \begin{bmatrix}
0 & 1 \\ -4 & 0
\end{bmatrix} x,
\end{align}
$u_t \in \R^2$ is a control input vector, $x_t \in \R^2$ is a state vector, and $t \in [0,\infty)$. 
The origin of the system can be stabilized by just the addition of noises. For example, if we employ $u$ as a Gaussian white noise, then \eqref{eq:simple1} becomes\footnote{More precisely, we design $u=\dot{w}^{WZ}$, where $w^{WZ} \in \R^2$ is defined in \dref{def:awp}; then we obtain \eqref{eq:sys-mcawp} by the limiting operation as with \tref{the:wong}. }
\begin{align}\label{eq:sys-mcawp}
dx_t &= f(x_t) dt + g_1(x_t) \circ dw_t[1] + g_2(x_t) \circ dw_t[2] \nonumber \\
%\end{align}
%\begin{align}
	&= -\!\begin{bmatrix} 9 & 0 \\ 0 & 1 \end{bmatrix} \! dt \!+ \!\begin{bmatrix} 0 & 0 \\  1 & 0 \end{bmatrix} \! x_t dw_t[1] \!+ \!\begin{bmatrix} 0 & 1 \\ -4 & 0 \end{bmatrix}\! x_t dw_t[2].
\end{align}
Analyzing the system using the basic results of stochastic Lyapunov stability theory, all the sample paths of this system converges on the origin with probability one\footnote{The origin is globally asymptotically stable in probability because $V_1(x)=x^T x$ is a global stochastic Lyapunov function; briefly speaking, $(\mathcal{L}V_1)(x)$---as the same form as \eqref{eq:lv-wz} below---is equal to $-(1/2)x^T x$; it is negative definite for all $x \in \R^2$; see the references on stochastic stability, for example \cite{arnold1990,florchinger1993,khasminskii2012,kushner}.}. Fig.~\ref{fig:mcawp-state} shows an example of the sample paths.

\begin{figure}[tbp]
\centering

\begin{minipage}{\columnwidth}
\centering
\includegraphics[width=6cm]{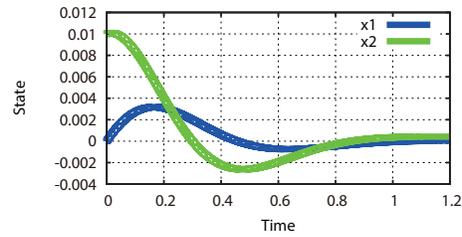}\\
(a)~$\eta=1$
\end{minipage}

\begin{minipage}{\columnwidth}
\centering
\includegraphics[width=6cm]{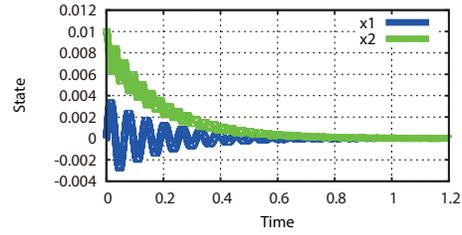}\\
(b)~$\eta=10$
\end{minipage}

\begin{minipage}{\columnwidth}
\centering
\includegraphics[width=6cm]{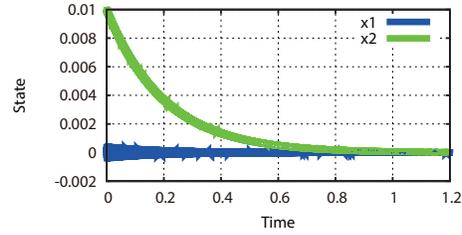}\\
(c)~$\eta=100$
\end{minipage}

\caption{Paths of \eqref{eq:simple1} with $u=\dot{\tilde{u}}^D(\eta)$ , $b_1=3$ and $b_2=4$.}
\label{fig:decnoise-state}
\end{figure}

Emphasizing again that the unstable origin becomes the stable origin by just adding the Gaussian white noises---roughly speaking, the noises are ``derivatives'' of Wiener processes. There are two candidates for the cause of the stabilization result: the randomness of the Wiener process and the unboundedness of the variations of the processes. In this paper, we claim that the main cause is the unboundedness, generally not the randomness because there are deterministic control inputs $u=\dot{\tilde{u}}^D(\eta)$, where
\begin{align}\label{eq:inp-ex1}
\tilde{u}^D_t[1](\eta) = b_1 \frac{\cos(\eta^2 t)-1}{\eta},\ \tilde{u}^D_t[2](\eta) = b_2 \frac{\sin(\eta^2 t)}{\eta}
\end{align}
with $\eta =1,2,\ldots$, deriving a stabilization result similar to the above example; see Fig.~\ref{fig:decnoise-state} as the circumstance evidence. As $\eta$ gets larger, the trajectories seem to close with a particular smooth trajectory that converges on the origin; to investigate the true identity of the ``pseudo-smooth trajectory'', we need consider the situation of $\eta \to \infty$. However, if $\eta \to \infty$, the derivatives of $\tilde{u}^D_t[1]$ and $\tilde{u}^D_t[1]$ are impossible to be defined because they become signals having unbounded variations, as with Wiener processes.

Thus, to deal with ``deterministic unbounded-variation noises'' such as $\lim_{\eta \to \infty} \tilde{u}(\eta)$ as with ``stochastic unbounded-variation noises'' such as Wiener processes, a comprehensive representation method of deterministic and stochastic systems is needed. Fortunately, the recent works on rough path analysis \cite{lyons2002,lyons2007,friz2014} enable us to deal with such unbounded-variation noises in a unified way. Then, in \cite{nishimura2015micnon}, Lyapunov-like stability theory for rough path analysis, namely ASiR, has been proposed. In this paper, we develop the theory by providing the proofs for the related rough systems and making further discussions so that the analysis of {\it stabilization by noise} is more generalized by allowing deterministic unbounded-variation noises. Furthermore, we will achieve that deterministic unbounded-variation noises provide stability property tighter than stochastic unbounded-variation noises.

%
%\begin{remark}
%The similar problem formulation of stabilization by $\tilde{u}^D(\eta)$ has been already discussed in the approximation algorithms \cite{sussmann1991,liu1997}; however, the algorithm has not clarified the relationship of stability properties between stochastic and deterministic noises. \eot
%\end{remark}
%}

%\begin{remark}

%\end{remark}

%%%%%%%%%%
\section{Rough Path Analysis}\label{sec:roughpath}

 To achieve our system representations including both deterministic and stochastic unbounded-variation noises, we should consider state variables and their dynamics allowing such noises. Because we claim that {\it rough paths} are suited for our state variables, we briefly summarize the rough path analysis based on \cite{lyons2007} in this section.

%%%%%
\subsection{Rough Path}

 Here we start with the definition of $p$-variations that are necessary to achieve the notion of rough paths:

\begin{definition}[$p$-variation \cite{lyons2007}]\label{def:variation}
Let $p \ge 1$, a continuous path $x: \Delta_T \rightarrow \R^n$, and a sequence of time sub-intervals $D$ be considered. Then,
\begin{align}
||x||_{p,J} = \left[ \sup_{D \subset J} \sum_{k=0}^{N-1} |x_{t_k,t_{k+1}}|^p \right]^{1/p}
\end{align}
is said to be the $p$-variation of $x$ (on the interval $J$). Furthermore, if $||x||_{p,J} < \infty$, then $x$ is said to have finite $p$-variation (on the interval $J$). \eod
\end{definition}

 To simplify the discussion, we consider the situation of $p \in [2,3)$ in what follows\footnote{Of course, we can consider rough path analysis for any large $p \ge 1$; however,  considering $p \in [2,3)$ is sufficient to discuss the relationship between deterministic noises \eqref{eq:inp-ex1} and Wiener processes constructed by \tref{the:wong}.}; this assumption also reduces the complexity of definitions hereafter. The next definition describes the predecessor of rough paths: 

\begin{definition}[multiplicative functional \cite{lyons2007}]\label{def:multiplicative}
Let $X: \Delta_T \rightarrow T^{2}(\R^n)$ be a continuous map satisfying
\begin{align}
X_{s,t} &= (X_{s,t}^0,X_{s,t}^1,X_{s,t}^{2}) 
\end{align}
for each $(s,t) \in \Delta_T$, where $X_{s,t}^0 \in \R$, $X_{s,t}^1 \in \R^n$, and $X_{s,t}^2 \in \R^n \otimes \R^n$. The function $X$ is called a multiplicative functional (of degree 2 in $\R^n$) if $X_{s,t}^0=1$ for all $(s,t) \in \Delta_T$ and
\begin{align}
X_{s,\tau} \otimes X_{\tau,t} = X_{s,t},\ \forall \tau \in [s,t].
\end{align}
\eod
\end{definition}

Now we are ready to define rough paths as follows:

\begin{definition}[rough path \cite{lyons2007}]\label{def:roughpath}
Let $X$, $\tilde{X}(1),\ldots,\tilde{X}(\eta): \Delta_T \rightarrow T^{2}(\R^n)$ be continuous maps. Then,
\begin{enumerate}
\item $X_{s,t}$ is said to be a {\it rough path} (of degree $p$ in $\R^n$) if it is a multiplicative functional with $X^1$ having finite $p$-variation;
\item $\tilde{X}(1)_{s,t},\ldots,\tilde{X}(\eta)_{s,t}$ are said to be {\it smooth rough paths} if they are multiplicative functionals with $\tilde{X}(1)^1,\ldots,\tilde{X}(\eta)^1$ having finite $1$-variations, respectively; and
\item $X_{s,t}$ is said to be a {\it geometric rough path} if it satisfies $d_p(\tilde{X}(\eta)_{s,t}, X_{s,t}) \rightarrow 0$ as $\eta \rightarrow \infty$, where\footnote{To be precise, we have to define $p$-variation topology in Definition 3.12 of \cite{lyons2007} after the definition of control functions, which is a completely different notion from control inputs in this paper, in Definition 1.9 of \cite{lyons2007}. While they are necessary to provide geometric rough paths and prove \tref{the:lyons} below, we abbreviate them to simplify the structure of this paper.}
\begin{align}
d_p(X,Y) &:= \max_{1 \le i \le 2} \sup_{D \subset J} \left( \sum_{k=1}^N||X^i_{t_{k-1},t_k}-Y^i_{t_{k-1},t_k}||^{\frac{p}{i}} \right)^{\frac{1}{p}}. \label{eq:distance}
\end{align}
\end{enumerate}
Furthermore, an element $X^j_{s,t}$ for $j = 0,1,2$ is said to be the $j$-th level path of $X_{s,t}$. \eod
\end{definition}

Note that, we mainly consider the first level paths of geometric rough paths as state variables; rough paths and smooth rough paths are introduced for the definition of geometric rough paths; and the second level paths are needed for describing the dynamics of the first level paths; see also the discussions in the rest of this section.

%\begin{remark}
%In \cite{lyons2007}, rough paths are defined in Banach spaces while we only consider $\R^n$. The main reason is that we will construct stability analysis carefully. Of course, the extension of spaces is one of our exciting future works. \eot
%\end{remark}

\begin{remark}
The second level paths of rough paths generally have flexibility of the definitions; for example, if $X^1_{s,t} = w(t)-w(s)$, $X^2_{s,t}$ can be defined by using whether It\^o integral $X^2_{s,t}=\int_s^t w(\tau) \otimes dw(\tau)$ or Stratonovich integral $X^2_{s,t}=\int_s^t w(\tau) \otimes \circ dw(\tau)$. However, if $X^2_{s,t}$ are created by It\^o integrals, $X$ is not a geometric rough path because there is no smooth rough path satisfying $d_p(\tilde{X}(\eta)_{s,t}, X_{s,t}) \rightarrow 0$ as $\eta \rightarrow \infty$, see Sec.~3.3 in \cite{lyons2007}. In contrast, if $X^2$ is constructed by Stratonovich integrals, $X$ is a geometric rough path because the approximation theorems \cite{ikeda1989,mcshane1972,sussmann1991}, typified in Wong and Zakai \cite{wong1965}, imply that the existence of $\tilde{X}(\eta)$ such that $d_p(\tilde{X}(\eta)_{s,t}, X_{s,t}) \rightarrow 0$ as $\eta \rightarrow \infty$; see Sec.~3.4 in \cite{lyons2007} or \cite{friz2009}. %This means that, if we consider It\^o-type stochastic differential equations as the original dynamics of target systems, we have to transform it into Stratonovich-type ones. \eot
\end{remark}

%%%%%
\subsection{Integration Along Rough Paths}\label{subsec:roughintegral}

 Because we will consider (the first level paths of geometric) rough paths as state variables of dynamical systems, we should investigate the behavior of rough paths concerned with time. This aim needs differential equations for rough paths; however, the equations are incapable of being defined directly because the first level paths of rough paths generally have unbounded $1$-variations; that is, they are undifferentiable almost everywhere, such as Wiener processes. Therefore, we should consider to derive integrals along rough paths and their integral equations as with stochastic systems. According to \cite{lyons2007}, we show the definition of integrals along rough paths. For $s=t_0 \le \cdots \le t_k=t$, $(s,t) \in \Delta_T$ and a smooth function $h=(h_1,h_2,\ldots,h_q)$ with $h_1,h_2,\ldots,h_q:\R^q \rightarrow \R^q$, let us consider 

\begin{align}
&I_h (X,D)^l_{s,t} := \sum_{k=1}^N I_h(X)^l_{t_{k-1},t_k}
\end{align}
with $l=1,2$ and
\begin{align}\label{eq:roughintegral-base}
I_h (X)^1_{t_{k-1},t_k} := &h_j(X^1_{t_{k-1}})X^1_{t_{k-1},t_k} + (\nabla H)(X^1_{t_{k-1},t_k}), \\
I_h (X)^2_{t_{k-1},t_k} := &h(X^1_{t_{k-1}})\otimes h(X^1_{t_{k-1}}) \cdot X^2_{t_{k-1},t_k},
\end{align}
where %$\nabla H \in \R^n$ is defined by the following elements:
\begin{align}
\!\!\! (\nabla H)(X^1_{t_{k-1},t_k})[i] := \tr \left[ \pfrac{h_i}{X^1}(X^1_{t_{k-1}}) \cdot X^2_{t_{k-1},t_k} \right]
\end{align}
for $i=1,2,\ldots,q$. Then, we define the following:
\begin{definition}
Let $X_{s,t}:\Delta_T \rightarrow T^{2}(\R^n)$ be a geometric rough path and $h(X^1_\tau)$ be smooth enough. Then,
\begin{align}\label{eq:roughintegral}
\int_s^t h(X_\tau) dX^l_\tau := \lim_{|D_T|\rightarrow 0} I_h(X,D)^l_{s,t}
\end{align}
is said to be an integral of $h$ along $X^l$ for $l=1,2$. Further,
\begin{align}
\int_s^t h(X_\tau) dX_\tau := \left(1, \int_s^t h(X_\tau) dX^1_\tau ,\int_s^t h(X_\tau) dX^2_\tau \right)
\end{align}
is said to be an integral of $h$ along $X$.\eod
\end{definition}

\begin{remark}
The key point of the definition along rough paths is the usage of \eqref{eq:roughintegral-base}; $I_h(X)^1$ is considered including the second order Taylor polynomial of $h(X^1)$ if we consider $X^1 \in \R$, $X^1_{s,t}=X^1_t-X^1_s$ and $X^2=(X^1_t-X^1_s)^2/2$. Roughly speaking, the integration along rough paths is an extended notion of integrations by considering higher-order Taylor expansions. \eot
\end{remark}

%\begin{remark}\label{rem:integration}
%Strictly speaking, the integral in \eqref{eq:roughintegral} is derived based on the notion of {\it almost rough paths}; see Exercise 4.11 in \cite{lyons2007}. However, the explanation of almost rough paths is too long to place the paper. On the other hand, \cite{friz2014} defines {\it rough integrals} very similar to \eqref{eq:roughintegral}, and note that, they argue that the rough integral is a definition more general than the integral constructed from almost rough paths because of its acceptability of a wider class of rough signals. In this paper, we do not discuss the issue because we only consider the class of rough paths considered in \cite{lyons2007}. For more information, see Chapter 4 in \cite{lyons2007} and Chapter 4 in \cite{friz2014}. Of course, both the rough integrals and the integrals based on almost rough paths are capable of being defined for any large $p>0$. The reason for the restriction to $p \in [2,3)$ is only for simplicity; that is, the restriction is enough to discuss the relationship between deterministic noises \eqref{eq:inp-ex1} and (Wong-Zakai-type) artificial Wiener processes. \eot
%\end{remark}

%%%%%
\subsection{Rough Systems}\label{subsec:rde}

Through the definitions of rough paths and integration along rough paths, we eventually achieve representations of dynamics for rough paths. Let us consider a nonlinear system\footnote{If $u$ has finite $1$-variation, \eqref{eq:sys-base} is equal to $\dot{x}=g(x) \dot{u}$. }
\begin{align}\label{eq:sys-base}
x_{s,t}&= \sum_{k=1}^N g(x_{t_{k-1}}) u_{t_{k-1},t_k} ,
\end{align}
where $(s,t) \in \Delta_T$, $x:=(x[1],\ldots,x[n])^T: \Delta_T \rightarrow M_X = \R^n$ is a state vector, $u:=(u[0],u[1],\ldots,u[m])^T :\Delta_T \rightarrow M_U = \R^{m+1}$ is an input vector, $u_\tau[0] \equiv \tau \in [s,t]$, and $g =(g_0,g_1,\ldots,g_d)$ with $g_0,g_1,\ldots,g_d: M_X \rightarrow M_X$ are assumed to be smooth and locally Lipschitz. Furthermore, we only consider the combination of $g$ and $u$ providing a unique global solution to \eqref{eq:sys-base}\footnote{In \cite{lyons2007}, $g$ is assumed to be globally bounded for the existence and the uniqueness of global solutions. However, we consider $g$ satisfying local bounded conditions; the lessening of the conditions on $g$ is necessary to consider our problem formulation, for example, \eqref{eq:simple1} in \sref{sec:motivation}. The detailed discussion on more concrete assumptions on $g$ and $u$ will be described in the next section; see B1 and B2 below.}.

%Here we define rough systems.

%$X_{s,t}:\Delta_T \rightarrow T^{2}(M_X)$ and

%\begin{definition}[rough differential equation \cite{lyons2003,lyons2007}]
%Let us consider \eqref{eq:sys-base}. A geometric rough path $\bar{Z}(t) \in T^{2}(\R^n) \oplus T^{2}(\R^m)$ is a solution to the differential equation
%\begin{align}\label{eq:rde}
%d\bar{X}(t) = G(\bar{X}(t),\bar{U}(t)) d\bar{U}
%\end{align}
%if all the following conditions hold:
%\begin{align}\label{eq:rde-integral}
%\bar{Z}_{s,t} = \int_s^t \begin{pmatrix} I_m & 0 \\ G(\bar{Z}(\tau)) & 0 \end{pmatrix} d\bar{Z}(\tau)
%\end{align}
%and $\pi_{\R^m}(\bar{Z}) = \bar{U}$, where $\pi_V: V \oplus W \rightarrow V$ for any spaces $V$ and $W$. \eod
%\end{definition}

%To analyze the dynamics of $x$ based on the rough path $U$, we consider the following system representation:

Let a geometric rough path $U :\Delta_T \rightarrow T^{2}(M_U)$, a map $X:\Delta_T \rightarrow T^{2}(M_X)$, and a map $Z:\Delta_T \rightarrow T^{2}(M_U) \oplus T^{2}(M_X)$ satisfying $\pi_{M_U}(Z) = U$, $\pi_{M_X}(Z) = X$; i.e.,
\begin{align}
Z^1_{s,t} = \begin{bmatrix}
U^1_{s,t} \\ X^1_{s,t} \end{bmatrix},\ Z^2_{s,t} = \begin{bmatrix}
U^2_{s,t} & Y^{UX}_{s,t} \\
Y^{XU}_{s,t} & X^2_{s,t}
\end{bmatrix}
\end{align}
be considered for $s \le \tau \le t \in [0,T]$ and $X^2: \Delta_T \rightarrow \R^n \otimes \R^n$, $Y^{UX}: \Delta_T \rightarrow \R^m \otimes \R^n$, and $Y^{XU}: \Delta_T \rightarrow \R^n \otimes \R^m$.%; concrete examples of these elements are shown in the next section}.

\begin{definition}[rough system and rough solution]\label{def:rde}
%Let $M_U$ $ = \R^{m+1}$ and $M_X =\R^n$. 
An equation %for a smooth function $g: M_X \rightarrow M_X \otimes M_U $,
\begin{align}\label{eq:rde-integral}
Z_{s,t} = \int_s^t \begin{bmatrix} I_m & 0 \\ g(X^1_\tau) & 0 \end{bmatrix} dZ_\tau
\end{align}
and $Z_{s,t}$ are said to be a {\it whole rough system} and a {\it whole rough solution}, of \eqref{eq:sys-base}, respectively. Furthermore,
\begin{align}\label{eq:rde-extracted}
X^1_{s,t} = \left[\pi_{M_X}\left(\int_s^t \begin{bmatrix} I_m & 0 \\ g(X^1_\tau) & 0 \end{bmatrix} dZ_\tau\right)\right]^1
\end{align}
extracted from \eqref{eq:rde-integral} and $X^1$ are said to be a {\it rough system} and a {\it rough solution}, of \eqref{eq:sys-base}, respectively. \eod
\end{definition}

The following result is a simplified version of Lyons universal limit theorem:

\begin{theorem}[\cite{lyons2007}]\label{the:lyons}
Let $\tilde{u}(\eta)$ for $\eta =1,2,\ldots$ have finite $1$-variations and $\tilde{U}(\eta)$ be a smooth rough path satisfying $\tilde{U}^1(\eta)_{s,t}=\tilde{u}_{s,t}(\eta)$. Let also $U_{s,t}$ be a geometric rough path satisfying $d_p(\tilde{U}_{s,t}(\eta),U) \rightarrow 0$ as $\eta \rightarrow \infty$ and the initial value $Z_s \in T^{2}(M_U) \oplus T^{\lfloor p \rfloor)}(M_X)$ be fixed. If \eqref{eq:sys-base} has a unique global solution with $u=\tilde{u}(\eta)$ for any $\eta$, the whole rough solution $Z_{s,t}$ of \eqref{eq:sys-base} uniquely exists and $Z_{s,t}$ is a geometric rough path. \eot
\end{theorem}

This result immediately implies that $X \in T^{2}(M_X)$ is also a geometric $p$-rough path of $x$. In this way, we obtain \eqref{eq:rde-extracted} as the dynamics of our state variable $X^1$; note that, \tref{the:lyons} implies that we should define the whole rough system \eqref{eq:rde-integral} because the dynamics of $X^1$ is directly affected by $U^2$; see also \rref{rem:itomap} below. This fact will be clearly articulated by deriving more specific representations in the next section.

%\begin{theorem}[\cite{lyons2007}]\label{the:lyons}
%Let $\tilde{u}(\eta)$ for $\eta =1,2,\ldots$ have finite $1$-variations and $\tilde{U}(\eta)$ be a smooth rough path satisfying $\tilde{U}^1(\eta)_{s,t}=\tilde{u}_{s,t}(\eta)$.} Let also $U_{s,t}$ be a geometric rough path satisfying $d_p(\tilde{U}_{s,t}(\eta),U) \rightarrow 0$ as $\eta \rightarrow \infty$} and the initial value $Z_s \in T^{2}(M_U) \oplus T^{\lfloor p \rfloor)}(M_X)$ be fixed. If \eqref{eq:sys-base} has a global solution with $u=\tilde{u}(\eta)$ for any $\eta$,} the whole rough solution $Z_{s,t}$ of \eqref{eq:sys-base} exists and $Z_{s,t}$ is a geometric rough path. Furthermore, if the global solution of \eqref{eq:sys-base} with $u=\tilde{u}(\eta)$ is unique, $Z_{s,t}$ is also unique. }\eot
%\end{theorem}
%
%This result immediately implies that $X \in T^{2}(M_X)$ is also a geometric $p$-rough path of $x$.
%%
%\begin{remark}\label{rem:localsolution}
%This section shows the basic definitions and results in \cite{lyons2007}, provided that the literature deals with $g$ satisfying global bounded conditions. Because we consider $g$ satisfying local bounded conditions, we need the complicated statements concerned with the existence and uniqueness of solutions in \tref{the:lyons}. The lessening of the conditions on $g$ is necessary to consider our problem formulation, for example, in \sref{sec:motivation}. The assumptions B1 and B2 are efficiently used for our main results. \eot
%\end{remark}
%}

\begin{remark}
Note that, in \cite{lyons2002,lyons2007}, a whole rough system is said to be {\it a rough differential equation} with the abbreviation of integral signs ``$\int$'' and $X$ is said to be a rough solution. \eot
\end{remark}

%\begin{remark}
%In \eqref{eq:sys-base}, we consider $g$ as $g(x,u)$ while the previous works \cite{lyons2002,lyons2007}] deal with $g(x)$. The extension does not cause any problem because we consider $Z = (U,X)$ in \eqref{eq:rde-integral} as the rough solution in either case. \eot
%\end{remark}

\begin{remark}\label{rem:itomap}
Despite focusing attention on the dynamics of $X^1$, we have to define the whole rough system \eqref{eq:rde-integral}. The reason is that $\int X dU$ is incapable of being defined if $X$ and $U$ have different ranges, i.e., $T^{2}(M_U) \neq T^{2}(M_X)$ while $\int Z dZ$ can be defined \cite{lyons2002,lyons2007}. Furthermore, The It\^o map of $X^1$, i.e., $U^1 \mapsto X^1$ (note that $U_t[0]=t$), is generally discontinuous while one of $Z$, i.e., $U \mapsto Z$, is continuous \cite{lyons2002,lyons2007}---the latter map is said to be It\^o-Lyons map \cite{friz2009}. In theory, this point is the true identity of ``the bridge'' between deterministic and stochastic differential equations. \eot
\end{remark}

%%%%%%%%%%%%%%%%%%%
\section{Stabilization by Unbounded Noises}\label{sec:main}

In this section, we investigate the effects of unbounded-variation noises---such as the limitation of $\tilde{u}(\eta)$ as $\eta \to \infty$ in \sref{sec:motivation} and Wiener processes---on stability properties for dynamical systems.

%%%%%
\subsection{System Representation}\label{subsec:canonical}
To begin with, we formulate rough systems more specific than the previous section.

In what follows, Let us consider the followings:
\begin{description}

\item[A0] $\tilde{u}(\eta)$ has finite $1$-variations for all $\eta =1,2,\ldots$;
\item[A1] $\tilde{U}(\eta): \Delta_\infty \rightarrow T^{2}(M_U)$ is a smooth rough path satisfying
\begin{align}
\hspace{-0.5cm}\tilde{U}_{s,t}^1(\eta) = \tilde{u}_{s,t}(\eta),\ \tilde{U}_{s,t}^2(\eta) = \int_s^t \tilde{u}_{s,\tau}(\eta) \otimes d\tilde{u}(\eta)_{\tau}
\end{align}
for all $\eta = 1,2,\ldots$; and
\item[A2] $U_{s,t}: \Delta_\infty \rightarrow T^{2}(M_U)$ is a geometric rough path such that $d_p(\tilde{U}(\eta)_{s,t},U_{s,t}) \rightarrow 0$ as $\eta \rightarrow \infty$.
\end{description}
%
%Furthermore, we assume that $\tilde{u}(\eta)$ and \eqref{eq:sys-base} satisfy either the following B1 or B2 depending on target systems:
%\begin{description}
%\item[B1] The signal $\tilde{u}(\eta)$ is piecewise linear for all time intervals. Furthermore, if $u_{t_{k-1},t_k}[j]=a_j$ with any constant $a_1,\ldots,a_m \in \R$ for any $j=1,2\ldots,m$ and for any time interval, \eqref{eq:sys-base} has a unique global solution.
%\item[B2] The signal $\tilde{u}(\eta)$ satisfies $\tilde{u}_{t_{k-1},t_k}(\eta)[j] \to 0$ as $\eta \rightarrow 0$ for any $j=1,2\ldots,m$ and for all time intervals. Furthermore, if $u_{t_{k-1},t_k}=0$ for any time interval, \eqref{eq:sys-base} has a unique global solution.
%\end{description}
%These assumptions are required for calculating \eqref{eq:distance}; see also \rref{rem:solution}. 
%}
Then, the rough system is derived as follows:
\begin{theorem}\label{the:rde}
Let A0--A2 be assumed and $u=\tilde{u}(\eta)$. Then, the rough system of \eqref{eq:sys-base} is given by
\begin{align}
X^1_{s,t} = &\int_s^t \sum_{j=0}^m g_j(Z^1_\tau) dU^1_\tau[j] + \int_s^t  \sum_{j=0}^m \sum_{k=0}^m \pfrac{g_j}{X^1}(X^1_\tau) g_k(X^1_\tau) dU^2_\tau[k,j] \label{eq:base-rde}
\end{align} 
as long as it has a unique global solution. \eot
\end{theorem}

The proof is shown in \sref{sec:rde}; note that, although the assumptions A0--A2 have already appeared in Section 3 of \cite{lyons2007}, the concrete system representation shown in this theorem was not provided. The rough paths satisfying A0 are said to be {\it canonical rough paths} \cite{lyons2007}, and they are used for dealing with Wiener processes in rough path analysis; see also \ssref{subsec:relation}. In this paper, we also consider A0 for constructing ASiR properties.

\begin{remark}
Emphasizing again that, not only the rough solution $X^1$ is uniquely derived but also the rough system \eqref{eq:base-rde} includes the second level variations: there exist the second term of the right-hand side of \eqref{eq:base-rde}; of course, if $u$ has finite $1$-variation, the terms vanish and \eqref{eq:base-rde} is equal to the original system \eqref{eq:sys-base} as usual sense of Stieltjes integrals. Furthermore, if $u$ has finite $p$-variation with $p \ge 3$, we add terms of the third level variations to the right-hand side of \eqref{eq:base-rde} based on \eqref{eq:rde-integral}. \eot
\end{remark} %To simplify the discussion, we only consider $p \in [2,3)$ in what follows.

\subsection{Lyapunov Stability for Rough Systems}\label{subsec:stability}

In this section, we consider stability properties for the origin of the rough system \eqref{eq:base-rde}; that is, we consider $g_0(0) \equiv 0$ and a control input $u$ satisfying A0--A2 and the following:
\begin{description}
\item[A3] $U_{s,t}$ makes the origin of \eqref{eq:base-rde} a unique equilibrium.
\end{description}

%%%%%
%\subsection{Stability in roughness and rough Lyapunov functions}
Here we define the notions of stability for the origins of rough systems. 

\begin{definition}[stability in roughness]\label{def:stability}
Let $U_{s,t}$ be fixed such that A0--A3 hold and $Z_s$ be fixed for an initial time $s \ge 0$. The origin of \eqref{eq:sys-base} with $u=\lim_{\eta \rightarrow \infty}\tilde{u}(\eta)$, or \eqref{eq:base-rde}, is said to be
\begin{description}
\item[D\ref{def:stability}-1] {\it (uniformly) stable in roughness} if for each $\varepsilon>0$, there is $\delta = \delta(\varepsilon)>0$ such that
\begin{align}
|X^1_s| < \delta \Rightarrow |X^1_t|<\varepsilon
\end{align}
holds for all $t \ge 0$, uniformly in $s \ge 0$;
\item[D\ref{def:stability}-2] {\it (uniformly) locally asymptotically stable in roughness (locally ASiR)} if it is stable in roughness and $\delta$ can be chosen such that
\begin{align}
|X^1_s| < \delta \Rightarrow \lim_{t\rightarrow \infty} X^1_t = 0
\end{align}
holds uniformly in $s \ge 0$;
\item[D\ref{def:stability}-3] {\it (uniformly) globally asymptotically stable in roughness (globally ASiR)} if it is stable in roughness and moreover
\begin{align}
\lim_{t\rightarrow \infty} X^1_t = 0
\end{align}
holds for all $X^1_s \in \R^n$, uniformly in $s \ge 0$. \eod
\end{description}
\end{definition}

This definition is the same as Lyapunov stability, local AS, and global AS for the origins of ordinary differential equations \cite{haddad,khalil}; hence, a question arises whether the related Lyapunov stability theory for \dref{def:stability} is also the same or not. To investigate it, we consider the dynamics of a sclar function $v:\R^n \to \R$ as a candidate of ``Lyapunov functions''; let us consider the following enhanced system:
\begin{align}\label{eq:sys-lyap-base}
\bar{x}_{s,t} = \sum_{k=1}^N \bar{g}(\bar{x}_{t_k-1}) u_{t_{k-1},t_k},
\end{align}
where $\bar{x}=(x[0],x[1],\ldots,x[n])^T$ and
\begin{align}
\bar{g}(x) = \begin{bmatrix}
	(\lie{g_0}{v})(x) & \ldots & (\lie{g_m}{v})(x)\\
	& g(x) &
	\end{bmatrix}.
\end{align}

Thus, we obtain the following:
\begin{lemma}\label{lem:lf}
Let A0--A2 be assumed. Then, we obtain 
%\begin{align}
%&V^1_{s,t} =  \int_s^t \sum_{j=1}^m \pfrac{V^1}{X^1}(X^1_\tau) g_j (X^1_\tau) dU^1_\tau[j] \nonumber \\
%	&+ \int_s^t\sum_{j=0}^m \sum_{k=0}^m \sum_{l=1}^n \left[\pfrac{}{X^1[l]}\pfrac{V^1}{X^1}g_j \right] (X^1_\tau) g_k [l](X^1_\tau) dU^2_\tau[k,j] \label{eq:sys-lyap},
%\end{align}
\begin{align}
&V^1_{s,t} =  \int_s^t \sum_{j=1}^m (\lie{g_j}{V^1})(X^1_\tau) dU^1_\tau[j] + \int_s^t \sum_{j=0}^m \sum_{k=0}^m (\lie{g_k}{\lie{g_j}{V^1}})(X^1_\tau) dU^2_\tau[k,j] \label{eq:sys-lyap},
\end{align}
where $V^1 := \bar{X}^1[0]$ is said to be the fist-level path of a rough path $V$, as long as $\bar{X}^1$ exists as a unique global rough solution of \eqref{eq:sys-lyap-base}. \eot
\end{lemma}

The lemma is proved as with \tref{the:rde}; because the form of \eqref{eq:sys-lyap-base} is the same as \eqref{eq:sys-base}, the rough system is \eqref{eq:rde-integral} with replacing $Z$ by $\bar{Z} = (U,\bar{X})$ and $g$ by $\bar{g}$, respectively. Thus, we obtain \eqref{eq:sys-lyap}.

Now we assume the following with $D$ as an open subset of $\R^n$ including the origin $x \equiv 0$:
\begin{description}
\item[C1] $v: D \rightarrow \R$ is smooth for all $x \in D$ and there exist continuous, positive definite, and proper functions $W_1,W_2: D \rightarrow \R$ such that
\begin{align}
W_1(x_t) \le v(x_t) \le W_2(x_t)
\end{align}
for all $t \ge 0$ and $x \in D$.
\end{description}

Then, we obtain the following result:

\begin{theorem}\label{the:ode}
Let A0--A3 and C1 be assumed; let also \eqref{eq:sys-base} and \eqref{eq:sys-lyap} be considered with $X_s$ being fixed for an initial time $s \ge 0$. Moreover, assume that there exists a function $\mathcal{D}V^1:\R^n \rightarrow \R$ such that $V^1_{s,t} = \int_s^t (\mathcal{D}V^1)(X^1_\tau) d\tau$ holds in $D$. Then, the following results are obtained:
\begin{description}
\item[C\ref{the:ode}-1] if $(\mathcal{D}V^1)(X^1) \le 0$ for all $X^1 \in D$ and $t \ge 0$, the origin of \eqref{eq:sys-base} is stable in roughness;
\item[C\ref{the:ode}-2] if there exists a continuous positive definite function $W_3:D \rightarrow \R$ such that $(\mathcal{D}V^1)(X^1_t)=-W_3(X^1_t)$ holds for all $X^1 \in D$ and $t \ge 0$, the origin of \eqref{eq:sys-base} is locally ASiR;
\item[C\ref{the:ode}-3] if C1 and C\ref{the:ode}-2 are all satisfied for $D = \R^n$, the origin of \eqref{eq:sys-base} is globally ASiR. \eot
\end{description}
\end{theorem}

The proof is the same as one for the original Lyapunov stability, for example, Theorems 4.1 and 4.2 in \cite{khalil} because $V^1_{s,t} = \int_s^t (\mathcal{D}V^1)(X^1_\tau) d\tau$ is the same condition as Lyapunov functions for ordinary differential equations. Thus, we obtain stability properties for rough systems; we conclude that the function $v$ of C1, or the first level path $V^1$ of the rough path $V$, is said to be a {\it rough Lyapunov function} (RLF), a (strict) local RLF, or a (strict) global RLF, if it satisfies C\ref{the:ode}-1, 2, or 3, respectively.

\begin{remark}
Assumption A3 is natural because we consider the stability of a unique equilibrium; however, this assumption does not ensure that the origin of \eqref{eq:sys-base} with $\tilde{u}(\eta)$ is an equilibrium when $\eta$ is finite. This point makes strange phenomena on the stability analysis of rough systems; see \ssref{subsec:case}. \eot
\end{remark}

%%%%%
\subsection{Relationship with stochastic stability}\label{subsec:relation}

The previous subsection shows that stability in roughness is the same notion as stability for ordinary differential equations, provided that the system are not described by ordinary differential equations but rough systems. On the other hand, rough systems include stochastic systems; hence, stability in roughness should also be the same as some notion of stochastic stability. In this subsection, we investigate what is such stochastic stability.

In this subsection, we assume that $\tilde{u}(\eta)$ and \eqref{eq:sys-base} satisfy the following:
\begin{description}
\item[B1] The signal $\tilde{u}(\eta)$ is piecewise linear for all time intervals. Furthermore, if $u_{t_{k-1},t_k}[j]=a_j$ with any constant $a_1,\ldots,a_m \in \R$ for any $j=1,2\ldots,m$ and for any time interval, \eqref{eq:sys-base} has a unique global solution.
\end{description}

\begin{lemma}\label{lem:infinitesimal}
Let \eqref{eq:sys-base} and \eqref{eq:sys-lyap} be considered. Let $u=w^{WZ}(\eta)$, where $w^{WZ}(\eta)$ is defined in \dref{def:awp}. If B1 holds and an $(n+1)$-dimensional stochastic system
\begin{align}\label{eq:sto-ori}
\bar{x}_{s,t} = \int_s^t \bar{g}_0(\bar{x}) d\tau + \sum_{j=1}^m \int_s^t \bar{g}_m(\bar{x}) \circ dw_\tau[j]
\end{align}
has an unique global solution, then we obtain
\begin{align}\label{eq:sto-rde}
V^1_{s,t} = \int_s^t (\mathcal{L}V^1)(X^1_\tau) d\tau + \sum_{j=1}^m \int_s^t (\lie{g_j}{V^1})(X^1_\tau) dU^1_\tau[j],
\end{align}
where
\begin{align}
\!\!\!(\mathcal{L}V^1)(X^1_t) = (\lie{g_0}{V^1})(X^1_t) + \sum_{j=1}^m (\lie{g_j}{\lie{g_j}{V^1}})(X^1_t) \label{eq:lv-wz}
\end{align}
\eot
\end{lemma}
The proof is based on the discussion of Section 3 in \cite{lyons2007}; see \sref{sec:infinitesimal} below.

This lemma immediately means that $\mathcal{L}V^1$ is the same as the infinitesimal operation of $V^1$ for Stratonovich-type stochastic differential equations driven by standard Wiener processes \cite{khasminskii2012}. Thus, we obtain the following:

\begin{proposition}\label{cor:wongzakai}
Let \eqref{eq:sys-base} and \eqref{eq:sys-lyap} be considered. Let also $\tilde{u}(\eta)$ be $w^{WZ}$ as with \lref{lem:infinitesimal}. If B1 holds and
\begin{align}
(\lie{g_j}{V^1})(X^1_t)=0\label{eq:con_uasas}
\end{align}
holds for all $j = 1,2,\ldots,m$ and for all $X^1 \in \R^n$ and $t \ge 0$, then we obtain \eqref{eq:lv-wz} and $\mathcal{L}V^1=\mathcal{D}V^1$. \eot
\end{proposition}

Because uniform almost sure asymptotic stability (UASAS) \cite{bardi2005} requires conditions as with \eqref{eq:con_uasas}, this corollary implies that UASAS is the same property as ASiR. It is not a surprising result because the UASAS is ``almost the same'' as AS \cite{nishimura2013sice}.

At the same time, in \cite{nishimura2013sice}, it is proved that the addition of any Wiener process never makes the non-AS-origin become UASAS. In contrast, we claim that deterministic unbounded-variation noises, such as \eqref{eq:inp-ex1}, are capable of stabilizing the origin in the sense of ASiR, despite the stability being the same as UASAS. The next subsection shows positive effects of deterministic unbounded-variation noises by comparing with stochastic unbounded-variation noises.

%On the other hand, $\tilde{u}(\eta)$ is a deterministic signal such as \eqref{eq:inp-ex1}, we can consider ASiR properties in the same manner as UASAS properties. However, deterministic signals are capable of stabilizing the non-AS origins in the sense of ASiR while no artificial Wiener process has possibility to do that in the sense of UASAS. The negative result concerned with UASAS is briefly shown by \tref{the:necessary} in \sref{sec:uasas}, and the positive result concerned with ASiR is discussed in the following section by calculating some simple examples.
\begin{remark}
If B1 holds, there exists a unique global solution for \eqref{eq:sys-base} under $u=w^{WZ}(\eta)$ with finite $\eta$, where $w^{WZ}(\eta)$ is ``piecewise linear approximate Wiener processes''; see \sref{sec:awp} below. Further, the global solution of \eqref{eq:sto-ori} is needed for the existence and uniqueness of \eqref{eq:sto-rde}. The existence of the both unique global solutions enable us to measure $d_p(\tilde{\bar{X}}(\eta),\bar{X})$ as $\eta \to \infty$; thus, we are capable of using \tref{the:lyons}. Note that, B1 allows us to consider bilinear systems such as \eqref{eq:simple1} because, if all the control inputs $u[1],\ldots,u[m]$ are nothing but step inputs, then \eqref{eq:simple1} becomes a linear system; that is, the system never has any finite escape time for any initial state. Of course, if we employ more complicated construction procedure for Wiener processes, B1 is insufficient to ensure the existence and uniqueness of the solutions. \eot
\end{remark}

\begin{remark}
As stated in \sref{sec:awp}, there are other schemes for constructing Wiener processes not satisfying B1. Nevertheless, the results in \cite{nishimura2013sice} imply that no Wiener process make the non-AS-origin become UASAS; that is, as long as considering the negative result, it is enough to consider Wiener processes created in \sref{sec:awp}. \eot
\end{remark}

\subsection{Stabilization by Deterministic Unbounded Noises}\label{subsec:case}

Here we describe stabilization results by the addition of control inputs having unbounded variations with reconsidering the discussion on \sref{sec:motivation}. In this subsection, we assume that $\tilde{u}(\eta)$ and \eqref{eq:sys-base} satisfy the following:
\begin{description}
\item[B2] The signal $\tilde{u}(\eta)$ satisfies $\tilde{u}_{t_{k-1},t_k}(\eta)[j] \to 0$ as $\eta \rightarrow 0$ for any $j=1,2\ldots,m$ and for all time intervals. Furthermore, if $u_{t_{k-1},t_k}=0$ for any time interval, \eqref{eq:sys-base} has a unique global solution.
\end{description}

The following theorem clarifies ``the additional terms'' generated by the deterministic unbounded-variation noises \eqref{eq:inp-ex1}:

\begin{theorem}\label{the:example}
Let us consider \eqref{eq:sys-base} with $m=2$ and \eqref{eq:inp-ex1}. Let also assume that B2 holds and the integral form of an $n$-dimensional ordinary differential equation
\begin{align}
x_{s,t}& = \int_s^t \Bigl[ g_0(x_\tau) + \frac{b_1b_2}{2} \Bigl\{\pfrac{g_2}{x}(x_\tau) g_1(x_\tau)- \pfrac{g_1}{x}(x_\tau) g_2(x_\tau) \Bigr\} \Bigr]  d\tau \label{eq:sys-ex1}
\end{align} 
has a global solution. Then, the rough system is derived by substituting $x=X^1$ into \eqref{eq:sys-ex1}. \eot
\end{theorem}
The proof is shown in \sref{sec:example} below.

\begin{remark}
If B2 holds, there exist unique global solutions for \eqref{eq:sys-base} under $u=\tilde{u}(\eta)$ with finite $\eta$. For example, let us consider $n=1$, $m=1$ and $g(x)=(0,x^2)$; then we have $\dot{x}=x^2 \dot{\tilde{u}}(\eta)$; that is, the solution is $x_t=x_s/(1-x_s \tilde{u}_{s,t}(\eta))$. This implies that there exists a solution for all time $t \in[0,\infty)$ if $|\tilde{u}_{s,t}(\eta)| < |x_s|$ holds; this inequality always holds with sufficiently-large $\eta$ because B2 implies that the amplitudes of $\tilde{u}(\eta)$ comes close to zero as $\eta$ increases. This implies that, there exists $\eta'=\eta -a $ with some positive integer $a$ such that $\tilde{u}(\eta'+a)$ ensures global solutions for all $\eta'=1,2,\ldots$. On the other hand, the global solution to \eqref{eq:sys-ex1} is needed for the existence and uniqueness of the related rough system. Note that, the existence of global solution to \eqref{eq:sys-ex1} is ensured if $g$ is locally Lipschitz and there exists a global Lyapunov function; if finite escape time arises for some initial state, it contradicts the existence of a global Lyapunov function. In this way, we can measure $d_p(\tilde{X}(\eta),X)$ as $\eta \to \infty$ to employ \tref{the:lyons}. \eot
\end{remark}

%%%%%
\subsection{Discussion With Case Study}

\begin{figure}[tbp]
%\centerline{\includegraphics[width=7cm]{sussmann_w11.eps}}
\centerline{\includegraphics[width=6cm]{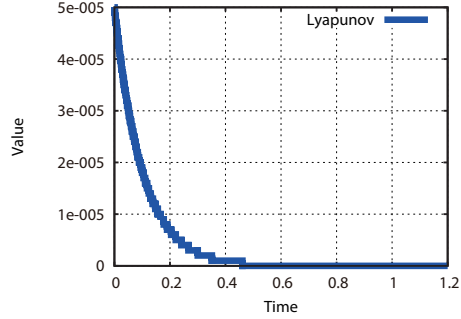}}
\caption{A path of $V^1$ with $u=\dot{\tilde{u}}^D(\eta)$, $\eta=100$, $b_1=3$ and $b_2=4$.}
\label{fig:decnoise-lyap}
\end{figure}

Here we reconsider the motivational example of deterministic noises in \sref{sec:motivation}. Using \tref{the:example}, we obtain 
\begin{align}
X^1_{s,t} = \frac12\int_s^t 
	 \begin{bmatrix}
	-14+b_1b_2 & 0 \\ 0 & 2-b_1b_2
	\end{bmatrix}
X^1_\tau d\tau
\end{align}
as the rough system of \eqref{eq:simple1} with $u=\dot{\tilde{u}}^D(\eta)$. Therefore, if $b_1b_2 \in (2,14)$, the origin is obviously globally ASiR because $V^1(X^1)=(1/2)(X^1)^T X^1$ immediately satisfies the condition C\ref{the:ode}-3. Fig.~\ref{fig:decnoise-lyap} describes the trajectory of this $V^1$ for the confirmation.

The answer to the motivational example shows that deterministic unbounded-variation noises are capable of making the non-AS-origin become the ASiR-origin. This implies that, the example is the case that stochastic and deterministic unbounded-variation noises both available for stabilization by noise; however, the results of \cite{nishimura2013sice} imply that stochastic unbounded noises do not help the non-AS-origin become the UASAS-origin. Thus, deterministic unbounded noises have possibility to achieve the stabilization results stronger than stochastic unbounded noises. In this subsection, we consider the reason by using a simpler example.

%Of course, according to \ssref{subsec:relation}, we conclude that the ASiR property ensured by deterministic noises is stronger than ASiP property ensured by stochastic noises; at the same time, deterministic noises are capable of simplifying the strategy of stabilization by noise by using \corref{cor:example}. Hence, in this subsection, we consider a simpler example to show another contribution of deterministic noises against stochastic ones. 

Considering $n=1$, $b_1=b_2=1$, and 
%the elements of the first level paths are calculated as
%\begin{align}
%\int_s^t dU^1_\tau[1] = \int_s^t dU^1_\tau[2] = \int_s^t dX^1_\tau = 0;
%\end{align}
%however, ones of second level paths are calculated as
%\begin{align}
%&\int_s^t dU^2_\tau[1,1] = \int_s^t dU^2_\tau[2,2]= \int_s^t dX^2_\tau = 0, \\
%&\int_s^t dU^2_\tau[1,2] = -\int_s^t dU^2_\tau[2,1]= \frac12 (t-s), \\
%&\int_s^t dC(U,X)_\tau[1,1] = - \int_s^t dC(X,U)_\tau[1,1] \nonumber \\
%&\hspace{2.9cm} = \frac12 \int_s^t g_{21}(Z_\tau)d\tau, \\
%&\int_s^t dC(X,U)_\tau[1,2] = - \int_s^t dC(U,X)_\tau[2,1] \nonumber \\
%&\hspace{2.9cm}= \frac12 \int_s^t g_{11}(Z_\tau)d\tau.
%\end{align}
%
%The calculation implies that, if 
\begin{align}
&g_0=0,\ g_1(x) = 1,\ g_2(x) = -x^2,\label{eq:ex2}
\end{align}
the rough system \eqref{eq:sys-ex1} is calculated as
\begin{align}\label{eq:sys-ex1-final}
&X^1_{s,t} = -\int_s^t X^1_\tau d\tau
\end{align}
while $\int_s^t dU^1_\tau = 0$. Obviously, the origin is globally ASiP.

%As a circumstance evidence for the rough system \eqref{eq:sys-ex1} to represent the dynamics of the system, we compare \eqref{eq:sys-base} with \eqref{eq:sys-ex1} via numerical calculations by using the most basic Euler scheme. 

Of course, the trajectories vibrate because $\tilde{u}^D_1(\eta)$ and $\tilde{u}^D_2(\eta)$ are the deterministic noises. However, the origin of the related rough system is globally ASiR. More clearly, if $\eta$ is finite, the system is represented by
\begin{align}\label{eq:sys-base-ex1}
x_{s,t} = \int_s^t \dot{\tilde{u}}^D_\tau[1](\eta) d\tau - \int_s^t x^2_\tau \dot{\tilde{u}}^D_\tau[2](\eta) d\tau;
\end{align}
that is, as long as $\eta$ is finite, the origin is never an equilibrium; of course, it is not globally AS. Nevertheless, the rough system \eqref{eq:sys-ex1-final} implies that the origin is globally ASiR. The trick is made by the calculation results $U^2 \neq 0$. 
%Note that, because we consider control inputs having bounded $p$-variations with $p \in [2,3)$, that is $p \neq 1$, the usual ordinary differential equations are incapable of being used; hence \eqref{eq:sys-base-ex1} is nothing but our picture for the comparison of rough systems with systems represented by ordinary differential equations. 
Thus, the representation of rough systems enlarges dynamical systems that we can consider.

The above stabilization result is due to the form of \tref{the:example}; for comparison, if $u=\dot{w}^{WZ}$ is chosen for \eqref{eq:sys-base} with \eqref{eq:ex2}, the resulting stochastic system is
\begin{align}
\label{eq:ex2-comp}x_{s,t} &= \int_s^t \circ dw_\tau[1] - \int_s^t x^2_\tau \circ dw_\tau[2],\\
\label{eq:ex2-comp-ito}	&=\int_s^t dw_\tau[1] - \int_s^t x^2_\tau dw_\tau[2] + \int_s^t x^3_\tau d\tau
\end{align}
if a solution exists\footnote{Note that, the existence of the solution to \eqref{eq:ex2-comp} is not always ensured.}.
Because the origin is not an equilibrium due to the form of the first term of the right-hand side of \eqref{eq:ex2-comp-ito}, it is not stable in probability. That is, the addition of Wiener processes makes the resulting stochastic systems have diffusion terms, namely the first and second terms of the right-hand side of \eqref{eq:ex2-comp-ito}. On the other hand, the addition of deterministic noises \eqref{eq:inp-ex1} does not cause such terms in rough systems, see again \eqref{eq:sys-ex1}.
%Furthermore, if the origin of the rough system of \eqref{eq:sys-base} is not asymptotically stable as $u_1=u_2=0$, it never becomes almost sure asymptotically stable as long as we choose control inputs including Wiener processes \cite{nishimura2013sice}]. In contrast, it becomes globally asymptotically stable if we design $u_1=u_1^\eta$ and $u_2=u_2^\eta$ as shown above\footnote{As shown in the another paper \cite{nishimura2015micnonb}], the ensured stability is the same as global almost sure asymptotic stability.}. 
This is the true identity of the advantage of deterministic unbounded-variation noises $\lim_{\eta \to \infty}\tilde{u}^D(\eta)$ over stochastic unbounded-variation noises $w$.

%%%%%
%\subsection{Stabilization by noise for chained systems}
%Finally, we clearly articulate the contribution of ASiR properties by applying a stabilization problem of a chained system with the use of the strategy of stabilization by noise.
%
%
%Consider a chained system; that is, \eqref{eq:sys-base} with $n=3$, $m=2$, and $g(x)[1] = (1, 0, x[2])^T$ and $g(x)[2]=(0,1,0)^T$. This system is well known as a class of nonholonomic systems \cite{}; that is, the origin is impossible to be stabilized by using any continuous state-feedback. Therefore, we should design discontinuous state-feedbacks \cite{}, time-varying state-feedbacks \cite{}, or noise-based state-feedbacks \cite{nishimura2013cdc,nishimura2013iscie}. 
%
%Focusing attention on stabilization by noise, we should consider non-quasi forms of SLFs and analyze the influences of $f^{WZ}$ in \eqref{eq:wz} and $\mathcal{L}^{Ito}$ in \eqref{eq:ito} \cite{nishimura2013cdc,nishimura2013iscie}. In contrast, our rough system \eqref{eq:sys-base-rde} 

%%%%%%%%%%
\section{Concluding Remarks}\label{sec:conclusion}

In this paper, we provided the availability of ``deterministic unbounded-variation noises'' by describing rough systems and their Lyapunov stability properties. In the procedure, we also clarified that the noises are capable of ensuring asymptotic stability in roughness, which is generally a stronger property than asymptotic stability in probability. Furthermore, the noises sometimes stabilize the origin of rough systems while the original system with bounded noises---they are considered as approximations of deterministic unbounded-variation noises---does not have the origins in equilibrium. 

%Because the paper is the first proposal as a journal article, we could not find spaces for some important discussions on asymptotic stability. For example, 
Finally, we should notice that the analyzed issue is similar to the problem formulation of stabilization by the approximation algorithms \cite{sussmann1991,liu1997}. The algorithm has not clarified the relationship between deterministic and stochastic noises yet. The comparison should be important to confirm that our strategy is effective for the application of stabilization by noise to nonholonomic systems \cite{nishimura2013cdc}. This point will be solve by considering rough paths of higher-order degrees with some improvement of the formulations.

%The first issue has possibilities for rough path analysis to connect stabilization theory to a wider world. The second one is strongly associated with the control performance as with stochastic asymptotic stability \cite{nishimura2013sice}]. The third one should be considered for deriving necessary conditions for stability in probability \cite{khasminskii2012}] and a wider use of stabilization by noise \cite{nishimura2014mtns}]. The fourth one is considered significant in designing strategies of disturbance attenuation \cite{nishimura2014msc}]. And the last one is quite important for investigating the effects of stabilization by noise on nonholonomic systems in differentiable manifolds; especially, the addition of deterministic noises has negative results if the state space is non-Euclidean \cite{}]; thus, we have to integrate these results into a whole. The five points will be quite interesting issues in our future works.

%\begin{ack}
%Place acknowledgments here.
%\end{ack}

%\bibliographystyle{plain} 

%\bibliography{ifacconf}             % bib file to produce the bibliography
                                                     % with bibtex (preferred)

\appendix

%%%%%
\section{Wiener Processes Created by Wong and Zakai \cite{wong1965}}\label{sec:awp}

In this section, we  consider a creation method of Wiener processes introduced by Wong and Zakai \cite{wong1965}.

\begin{definition}[\cite{wong1965}]\label{def:awp}
Let $N=\eta$. Let also bring on sequences of discrete-time Wiener processes $\{w^D_{t_0}[j](\eta)$, $\ldots$, $w^D_{t_N}[j](\eta)\}$ for $j =1,2,\ldots,m$ and $t_k \in (s,t) \in \Delta_T$. For every $\tau \in [t_k,t_{k+1})$,
\begin{align}
\hspace{-0.2cm}w^{WZ}_\tau[j](\eta) := w^D_{t_k}[j](\eta)+ \dfrac{\tau-t_k}{t_{k+1}-t_{k}} {w^D}_{t_k,t_{k+1}}[j](\eta).
\end{align}
Then, $w^{WZ}[j](\eta)$ for $j =1,2,\ldots,m$ is said to be a {\it  (approximate) Wiener process}. \eod
\end{definition}

\begin{theorem}[\cite{wong1965}]\label{the:wong}
Let $m =1,2,\ldots$. If $u=\dot{w}^{WZ}(\eta)$, \eqref{eq:sys-base} becomes
\begin{align}\label{eq:sys-sto}
x_{s,t} &= \int_s^t g_0(x_\tau) d\tau + \sum_{j=1}^2 \int_s^t g_j(x_\tau) \circ dw_\tau[j] \nonumber \\
	&= \int_s^t f(x_\tau)d\tau + \sum_{j=1}^2 \int_s^t g_j(x_\tau) dw_\tau[j]
\end{align}
as $\eta \to \infty$ with probability one, where
\begin{align}\label{eq:wz}
f(X^1_t) := g_0(X^1_t) + \sum_{j=1}^m \frac12 \pfrac{g_j}{X^1}(X^1_t)g_j (X^1_t).
\end{align}
\eot
\end{theorem}

\begin{remark}
The above-mentioned Wiener processes are piecewise linear interpolations of discrete-time Wiener processes; that is, $w^{WZ}$ satisfies B1 while $\eta=N$ is finite. Note that, there are some creation methods of Wiener processes, see \cite{ikeda1989,mcshane1972,sussmann1991}. If an approximate Wiener process is constructed by nonlinear interpolation functions, then the derivative does not consist of step inputs; in this case, B1 does not hold. \eot
\end{remark}

\section{Proof of \tref{the:rde}}\label{sec:rde}

%As shown in \rref{rem:integration}, integrals along rough paths are necessary to be strictly-defined for deriving the theorem while the detail of the procedure can not be shown in this paper due to its length. Here, we bring the result as follows [Exercise 4.11 in \cite{lyons2007}]:

%\begin{lemma}\cite{lyons2007}]
%For a $p$-rough path $Z \in T^{2}(\R^{q})$ with $p \in [2,3)$ and a smooth function $f:\R^q \rightarrow \R^q \oplus \R^q$, the first level path of $\int_s^t f(Z(\tau)) dZ(\tau)$ is 
%\begin{align}
%&f(Z(s)) Z^1_{s,t} + (\nabla f)(Z(s)) Z^2_{s,t},
%\end{align}
%where the $i$-th element of the second term is equal to
%\begin{align}
%\sum_{j=1}^{q} \sum_{k=1}^{q} \pfrac{f_{ji}}{Z^1_k}(Z(s)) \int_s^t (Z^1_k)_{s,\tau} d{Z^1_j}(\tau)
%\end{align}
%for $i \in \N_1^q$. \eot
%\end{lemma}

%Throughout the proof, we consider $s=t_{k-1}$ and $\tau=t_{k}$ for any $k \in \N_1^N$ for simplicity.

Let $\tilde{Z}:\Delta_T \rightarrow T^{2}(M_U) \oplus T^{2}(M_X)$ be the whole rough solution to \eqref{eq:sys-base} with $\pi_{M_U}(\tilde{Z})=\tilde{U}(\eta)$ and $\pi_{M_X}(\tilde{Z})=\tilde{X}(\eta)$ for all $\eta =1,2,\ldots$.  In this section, let $s'= t_{k-1}$ and $t'=t_k$ for any $k =1,2,\ldots,N$ for simplisity.

\subsection{Deriving a geometric rough path $Z$.}
First, we calculate the elements of the rough path $\tilde{Z}_{s',t'}$.
% of $z=(u^T,x^T)=(z[0],z[1],z[2],\ldots,z[n+m])^T$; that is,
%\begin{align}
%Z^1_{s,\tau} = \begin{bmatrix}
%	U^1_{s,\tau} \\ X^1_{s,\tau}
%	\end{bmatrix},\ 
%Z^2_{s,\tau} = \begin{bmatrix}
%	U^2_{s,\tau} & C(U,X)_{s,\tau} \\
%	C(X,U)_{s,\tau} & X^2_{s,\tau}
%	\end{bmatrix},
%\end{align}
%where 
%\begin{align}
%&C(U,X)_{s,\tau} = U^1_{s,\nu} \otimes X^1_{\nu,\tau},\\
%&C(X,U)_{s,\tau} = X^1_{s,\nu} \otimes U^1_{\nu,\tau},%,\\
%%&Z^1_U:=(Z^1[0],\ldots,Z^1[m])^T, \\
%%&Z^1_X:=(Z^1[m+1],\ldots,Z^1[n+m])^T.
%\end{align}
%with $s < \nu < \tau$. 
The first level path is immediately obtained as $Z^1_{s',t'} = \lim_{\eta \rightarrow \infty} \tilde{Z}^1_{s',t'}$. The second level path is derived as follows. For $i,l=1,2,\ldots,n$ and $j =1,2,\ldots,m$, 

\begin{align}
Y^{UX}_{s',t'}&[j,i] = \lim_{\eta \rightarrow \infty} \int_{s'}^{t'} (\tilde{U}(\eta))^1_{s',\tau}[j] d(\tilde{X}(\eta))^1_\tau[i] \label{eq:rde-conv1} \\
	&= \lim_{\eta \rightarrow \infty} \sum_{k=0}^m g_k[i]((\tilde{X}(\eta))^1_{s'}) \int_{s'}^{t'} (\tilde{U}(\eta))^1_{s',\tau}[j] d(\tilde{U}(\eta))^1_\tau[k] \label{eq:rde-conv2} \\
	&= \lim_{\eta \rightarrow \infty} \sum_{k=0}^m g_k[i]((\tilde{X}(\eta))^1_{s'}) (\tilde{U}(\eta))^2_{s',t'}[j,k] \label{eq:rde-conv3}\\
	%&= \sum_{k=0}^m g_k[i]((\tilde{X}(\eta))^1_{s'}) \lim_{\eta \rightarrow \infty} (\tilde{U}(\eta))^2_{s',t'}[j,k] \label{eq:rde-conv4}\\
	&= \sum_{k=0}^m g_k[i] (X^1_{s'})U^2_{s',t'}[j,k]. \label{eq:rde-conv4}
\end{align}
Note that, the above integrals in \eqref{eq:rde-conv1}--\eqref{eq:rde-conv2} are all Stieltjes integrals because $\tilde{U}^1(\eta)$ and $\tilde{X}^1(\eta)$ are nothing but the smooth functions $\tilde{u}(\eta)$ and $\tilde{x}(\eta)$ that have finite $1$-variations, respectively, where $\tilde{u}(\eta)$ is assumed as A0, $\tilde{U}^1(\eta)$ is defined in A1, and $\tilde{x}(\eta)$ is a solution to the original system \eqref{eq:sys-base} with $u_{s,t}=\tilde{u}(\eta)_{s,t}$. Transforming \eqref{eq:rde-conv3} into \eqref{eq:rde-conv4} is achieved by A2 and the assumption of the existence of a unique global solution $\tilde{x}(\eta)$ for any $\eta>0$.
%as follows: as $\eta \rightarrow \infty$, $\tilde{u}(\eta)_{s,t} = (\tilde{U}(\eta))^1_{s,t} \rightarrow U^1_{s,t}$, where $U^1$ is the first level path of the geometric $p$-rough path $U$; that is, $|U^1_{s,t}[k]| < \infty$ without depending on $\eta$ for every $k = 0,1,\ldots,m$. This implies that, $g(X^1_{t_0})U^1_{t_0,t_1}$ is also bounded without depending on $\eta$ when an initial value $x_{t_0}$ is fixed and $\tilde{x}(\eta)$ exists as a unique global solution. Thus, $X^1_{t_1}=X^1_{t_0} + g(X^1_{t_0})U^1_{t_0,t_1}$ is bounded. In this way, we conclude that $g(X^1_{s'})$ for every $s' \in \Delta_T$ is bounded; this result immediately implies that \eqref{eq:rde-conv3} is equivalent to \eqref{eq:rde-conv4}.}

Similarly, we can calculate 
\begin{align}
&Y^{XU}_{s',t'}[j,i] = \sum_{k=0}^m g_k[i](X^1_{s'}) U^2_{s',t'}[k,j],\\
&X^2_{s',t'}[i,l] %&= \lim_{\eta \rightarrow \infty} \tilde{X}^1_{s,\tau}[i] \tilde{X}^1_{\tau,t}[j] \nonumber \\
	=\sum_{j,k=0}^m g_j[i](X^1_{s'}) g_j[l](X^1_{s'}) U^2_{s',t'}[j,k].
\end{align}

%For $i,j \in \N_1^n$ and $j \in \N_1^m$, we pick an element:
%\begin{align}
%C(U,X)_{s,\tau}[j,i] &= {Z^1_U}_{s,\tau}[j] {Z^1_X}_{s,\tau}[i] \nonumber \\
%	&= \sum_{k=0}^m g(Z_s)[i,k] {Z^1_U}_{s,\tau}[j] {Z^1_U}_{s,\tau}[k] \nonumber \\
%	&= \sum_{k=0}^m g(Z_s)[i,k] U^2_{s,\tau}[j,k].
%\end{align}
%Similarly, we can calculate 
%\begin{align}
%&C(X,U)_{s,\tau}[i,j] = \sum_{k=0}^m g(Z_s)[i,k] U^2_{s,\tau}[k,j]
%\end{align}
%Furthermore, we obtain
%\begin{align}
%X^2_{s,\tau}(i,j) &= {Z^1_X}_{s,\tau} {Z^1_X}_{s,\tau} \nonumber \\
%	&=\sum_{j,k=0}^m g(Z_s)[i,j] g(Z_s)[j,k] U^2_{s,\tau}[j,k].
%\end{align}
Thus, all the elements of $Z^2_{s',t'}$ are calculated.

\subsection{Deriving a rough system.}
Using the form of \eqref{eq:roughintegral-base} with
\begin{align}
h(z)=\begin{pmatrix}
I_m & 0 \\ g(z) & 0
\end{pmatrix},
\end{align}
$q=n+m+1$, and $z=(u[0],\ldots,u[m],x[1],\ldots,x[n])^T$, we obtain
\begin{align}
(h(Z^1_{s'}) Z^1_{s',t'})[i] = \sum_{j=0}^m g_j[i](Z_{s'}) U^1_{s',t'}[i]
\end{align}
and
\begin{align}
(\nabla H)(X^1_{t_{k-1},t_k})[i] = &\sum_{j=0}^{m} \sum_{k=0}^{m+n} \pfrac{g_j[i]}{Z^1[k]}(X^1_{s'}) Z^1_{s',t'}[k] U^1_{s',t'}[j] \nonumber \\
	=&\sum_{j=0}^{m} \sum_{l=1}^{n} \pfrac{g_j[i]}{X^1_l}(X^1_{s'}) X^1_{s,\tau}[l] U^1_{\tau,t}[j] \nonumber \\
	=&\sum_{j=0}^{m} \sum_{k=0}^{m} \sum_{l=1}^{n} \pfrac{g_j[i]}{X^1[l]}(X^1_{s'})g_k[l](X^1_{s'}) U^2_{s',t'}[k,j]
\end{align}
for $i={m+1},\ldots,{m+n}$. Thus, we obtain $I_h(Z)^1_{s',t'}$. These calculation results and the definition of the integral along rough paths in \ssref{subsec:roughintegral} implies that \eqref{eq:base-rde} holds.

%%%%
\section{Proof of \lref{lem:infinitesimal}}\label{sec:infinitesimal}
If A0 and A1 hold, all the elements of the second level path of $\tilde{U}(\eta)$ are transformed into
\begin{align}
\tilde{U}^2_{s,t}[j,k](\eta) = \frac12 \tilde{u}_{s,t}[j](\eta) \tilde{u}_{s,t}[k](\eta) + A_{s,t}[j,k],
\end{align}
where $j,k =0,1,\ldots,m$ and
\begin{align}
A_{s,t}[j,k] = &\frac12 \int_s^t \tilde{u}_{s,\tau}[j](\eta) d\tilde{u}_\tau[k](\eta) - \frac12 \int_s^t \tilde{u}_{s,\tau}[k](\eta) d\tilde{u}_\tau[j](\eta).
\end{align}

On the other hand, if  we design $\tilde{u}_{\tau}(\eta)=(\tau,w^{WZ}_\tau[1]$ $,\ldots,$ $w^{WZ}_\tau[m])^T$, we obtain
\begin{align}
&\lim_{\eta \rightarrow \infty} w^{WZ}_t(\eta) = (t,w_t[1],\ldots,w_t[m])^T,\\
&\lim_{\eta \rightarrow \infty} \int_s^t w^{WZ}_{s,\tau}[j](\eta) dw^{WZ}_\tau[j](\eta) = t-s,\ j=1,2,\ldots,m \\
&\lim_{\eta \rightarrow \infty} \int_s^t w^{WZ}_{s,\tau}[j](\eta) dw^{WZ}_\tau[k](\eta) = 0,\ k \neq j,\ j=k=0, \label{eq:crossterm}
\end{align}
by using It\^o's stochastic analysis. Therefore, we obtain
\begin{align}
&U^1_{s,t} = (t-s, w_{s,t}[1],\ldots,w_{s,t}[m])^T, \\
&U^2_{s,t}[j,j] = \frac12 (t-s), \ j = 1,2,\ldots,m \\
&U^2_{s,t}[j,k] = 0,\ k \neq j,\ j=k=0,
\end{align}
with $A_{s,t}[j,k]=0$ for all $j,k=1,2,\ldots,m$ by using \eqref{eq:crossterm}.  Then, substituting these results into \eqref{eq:sys-lyap}, we obtain  \eqref{eq:sto-rde}.

%%%%%
\section{Proof of \tref{the:example}}\label{sec:example}

%\subsubsection{Deriving a geometric rough path $U$.}
To begin with, we consider $u=(u[0],u[1],u[2])^T$, where $u_t[0]=t$ and $u[j]=\tilde{u}^D(\eta)[j]$ for $j=1,2$. 
%The calculation
%\begin{align}
%&\lim_{\eta \rightarrow \infty}u_{s,t}[0] = t-s < \infty, \label{eq:roughpath-u-1}\\
%&\lim_{\eta \rightarrow \infty} u_{s,t}[1] = \lim_{\eta \rightarrow \infty} u_{s,t}[2] = 0,\label{eq:roughpath-u-2}
%\end{align}
%imply that the rough path $U$ of $u$ is a geometric $p$-rough path with some $p>0$. 
As $\eta \to \infty$, $p$-variation of $u$ is infinite if $p \in (0,2)$ and finite if $p \ge 2$; that is,  A0 holds. Hence, we can consider A1, and then we obtain a geometric rough path $U_{s,t}$ satisfying A2.
%Furthermore, we use a good property of geometric rough paths:
%\begin{lemma}\cite{lyons2002,lyons2007}]\label{lem:geometric-asym}
%If A1 holds, then 
%\begin{align}
%U^2_{s,t}[i,j] + U^2_{s,t}[j,i] = U^1_{s,t}[i] U^1_{s,t}[j]
%\end{align}
%is satisfied. \eot
%\end{lemma}
%Using this lemma and \eqref{eq:roughpath-u-1}--\eqref{eq:roughpath-u-2}, we obtain
%Thus, we obtain
%\begin{align}
%U^1_{s,t}=\begin{bmatrix}
%t-s \\ 0 \\ 0
%\end{bmatrix},\ U^2_{s,t}=\begin{bmatrix}
%0 & 0 & 0 \\
%0 & 0 & U^2_{s,t}[1,2] \\
%0 & U^2_{s,t}[2,1] & 0
%\end{bmatrix},
%\end{align}
%where 

Next, we calculate elements of $U$:
\begin{align}
U^1_{s,t}[1] &= U^1_{s,t}[2] = U^2_{s,t}[0,0] = U^2_{s,t}[0,1] = U^2_{s,t}[0,2] \nonumber \\
	&= U^2_{s,t}[1,1] = U^2_{s,t}[2,2] = 0,\\
U^2_{s,t}[1,2] &= \lim_{\eta \rightarrow \infty} (\tilde{U}(\eta))^2_{s,t}[1,2] \nonumber \\
	&= \lim_{\eta \rightarrow \infty} \int_s^t \{\tilde{u}^D_\tau[1](\eta)-\tilde{u}^D_s[1](\eta) \} \dot{\tilde{u}}^D_\tau[2](\eta) d\tau \nonumber \\
	&= \lim_{\eta \rightarrow \infty} \int_s^t b_1b_2\{ \cos^2(\eta^2 \tau)-\cos(\eta^2s) \cos(\eta^2 \tau) \} d\tau \nonumber \\
%	&=\frac12 (t-s) +\frac12 \lim_{\eta \rightarrow \infty} \left[ \frac{\sin(2\eta^2 \tau)}{2\eta^2} \right]_s^t \nonumber \\
%	&\quad - \frac12 \lim_{\eta \rightarrow \infty} \cos(\eta^2 s) \left[ \frac{\sin \eta^2 \tau}{\eta^2} \right]_s^t \nonumber \\
	&=\frac{b_1b_2}{2} (t-s) = -U^2_{s,t}[2,1].
\end{align}
Thus, we obtain
\begin{align}\label{eq:roughpathu}
U_{s,t} = \left( 1, \begin{bmatrix}
t-s \\ 0 \\ 0
\end{bmatrix}, \frac{b_1b_2}{2} (t-s)\begin{bmatrix}
0 & 0 & 0 \\
0 & 0 & 1 \\ 
0 & -1 & 0
\end{bmatrix} \right).
\end{align}

Thus, we conclude that \eqref{eq:sys-ex1} holds by using \eqref{eq:roughpathu} and \tref{the:rde}. 

\end{document}